\documentclass[pdftex,superscriptaddress,aps,onecolumn]{revtex4}
\usepackage{multirow,enumerate,epsfig,graphics,amssymb,amsmath,subeqnarray,mathrsfs,color,xcolor}
\usepackage{color,epsfig,graphics,amssymb,amsmath,subeqnarray,graphicx,amsthm,subfigure,mathrsfs}
\usepackage[colorlinks=true, citecolor=red, linkcolor=blue ]{hyperref}

\newcommand{\change}[1]{{\color{black} #1}}
\newcommand{\dd}{\mathrm{d}}
\newcommand{\ee}{\mathrm{e}}\newcommand{\ci}{\mathrm{i}}

\newcommand{\nb}{\mathbf{n}}

\newcommand{\ub}{\mathbf{u}}
\newcommand{\eb}{\mathbf{e}}

\newcommand{\rb}{\mathbf{r}}
\newcommand{\Pe}{\mbox{Pe}}

\let\grad\nabla
\let\grad\nabla

\newcommand{\pard}[2]{\frac{\partial #1}{\partial #2}}

\makeatletter
\def\sgn{\mathop{\operator@font sgn}}
\makeatother
\DeclareMathAlphabet{\mathcal}{OMS}{cmsy}{m}{n}

\begin{document}
\title{Spontaneous onset of convection in a uniform phoretic channel}
\author{S\'ebastien Michelin}
\email{sebastien.michelin@ladhyx.polytechnique.fr}
\affiliation{LadHyX -- D\'epartement de M\'ecanique, Ecole Polytechnique -- CNRS, Institut Polytechnique de Paris, 91128 Palaiseau, France.}
\author{Simon Game}
\affiliation{Department of Mathematics, Imperial College, London SW7 2BZ, United Kingdom}
\author{Eric Lauga}
\affiliation{Department of Applied Mathematics and Theoretical Physics, University of Cambridge, Cambridge, CB3 0WA, United Kingdom}
\author{Eric Keaveny}
\affiliation{Department of Mathematics, Imperial College, London SW7 2BZ, United Kingdom}
\author{Demetrios Papageorgiou}
\affiliation{Department of Mathematics, Imperial College, London SW7 2BZ, United Kingdom}

\date{\today}
\begin{abstract}
Phoretic mechanisms, whereby gradients of chemical solutes induce surface-driven flows,  have recently been used to generate directed propulsion of patterned colloidal particles. When the chemical solutes diffuse slowly, an instability further provides active but isotropic particles with a route to self-propulsion by spontaneously breaking the symmetry of the solute distribution. Here we show theoretically that, in a mechanism analogous to B\'enard-Marangoni convection, phoretic phenomena can create spontaneous and self-sustained wall-driven \change{mixing} flows within a  straight, chemically-uniform active channel. \change{Such spontaneous flows do not result in any net pumping for a uniform channel but greatly modify the distribution of transport of the chemical solute.} The instability is predicted to occur for a solute P\'eclet number above a critical value and for a band of finite perturbation wavenumbers. We solve the perturbation problem analytically to characterize the instability, and use both steady and unsteady numerical  computations of the full nonlinear transport problem  to capture  the long-time coupled dynamics of the solute and flow within the channel.
\end{abstract}
\maketitle

\section{Introduction}
The rapid development of microfluidics   has motivated extensive research aiming at the precise control of micro-scale fluid flows~\cite{ho1998,stone2004,whitesides2006}. The standard approach to drive flows uses macroscopic mechanical forcing, namely a pressure difference imposed between inlet and outlet channels, which is sufficient to overcome viscous resistance over the entire length of the microchannel~\cite{squires2005}. Yet, another possibility resides in a local forcing of the flow directly at the channel wall, as realized, for example, in biological systems through the beating of active cilia anchored at the wall that generate a net fluid pumping which is critical to many biological functions~\cite{sleigh1988}.

Generating similar wall forcing synthetically by mimicking biological cilia is a difficult task, which requires complex assembly of flexible structures and yet another macroscopic driving 
(e.g.~magnetic)~\cite{babataheri2011}. Phoretic phenomena have  emerged as alternatives to micro-mechanical forcing able to generate a local forcing on a fluid flow near surfaces without relying on complex actuation. Instead, they exploit the emergence of surface slip flows resulting  from local physico-chemical gradients within the fluid phase above a rigid surface~\cite{anderson1989,sia2003,ajdari1995,ajdari2000}. These gradients can be that of a chemical  (diffusiophoresis), thermal  (thermophoresis) or electrical field (electrophoresis). While  phoretic flows have long been studied under  external macroscopic  gradients, they can also arise locally when the surface mobility is combined with a physico-chemical activity that provide the wall with the ability to change its immediate environment. This combination, often termed self-phoresis, has received much recent interest for self-propulsion applications~\cite{paxton2004,duan2015,moran2017}. Self-phoresis has also recently been considered as a potential alternative to macroscopically-actuated phoretic driving in microchannels~\cite{michelin2015b,yang2016,michelin2019b}.

Whether they are used to drive fluid within a microchannel or to propel a colloidal particle, phoretic flows require the presence of  physico-chemical gradients. To achieve phoretic transport, the system must therefore be able to break the directional symmetry of the  field responsible for the phoretic forcing.  
For self-propulsion three different routes have been identified for setting a single colloid into motion, namely (i) a chemical patterning of the surface~\cite{paxton2004}, (ii) a geometric asymmetry of the particle~\cite{michelin2015a,shklyaev2014},  (iii) an instability mechanism resulting from the nonlinear coupling of a solute dynamics to the phoretic flows when diffusion is slow~\cite{michelin2013c}.
The first two approaches are intrinsically associated with an asymmetric design of the system, and have already been explored for generating phoretic flows in microfluidic setups~\cite{michelin2015b,yang2015,yang2016,michelin2019b}.  The focus of our paper is on the \change{ability of instabilities to generate} spontaneous flows in phoretic microchannels. 
 The instability exploits the nonlinear coupling of physical chemistry and hydrodynamics through the convective transport of solute species by the phoretic flows, and provides isotropic systems (e.g.~a chemically-homogeneous spherical particle) with a spontaneous swimming velocity~\cite{michelin2013c}.  \change{The purpose of the present study is to determine whether such instability exists in a channel configuration and the characteristics of the flow and solute transport it can generate (e.g. pumping or mixing flow).}

We focus throughout the manuscript on diffusiophoresis where the physico-chemical field of interest is the concentration of a solute species consumed or produced at the active wall. Yet, the conclusions of this work are easily extended to other phoretic phenomena. In that context, the ``isotropic'' system consists in the most typical microfluidic setting, namely a rectilinear microfluidic channel with chemically-homogeneous walls. An active wall 
(e.g.~releasing a chemical solute which is absorbed at the opposite wall, transported or degenerated within the channel) generates an excess solute concentration in its immediate vicinity and thus, a normal solute gradient. In a purely isotropic setting, the solute concentration is homogeneous in the streamwise direction, thus generating no slip forcing at the wall and no flow within the channel. 

\begin{figure*}
\begin{center}
\includegraphics[width=.9\textwidth]{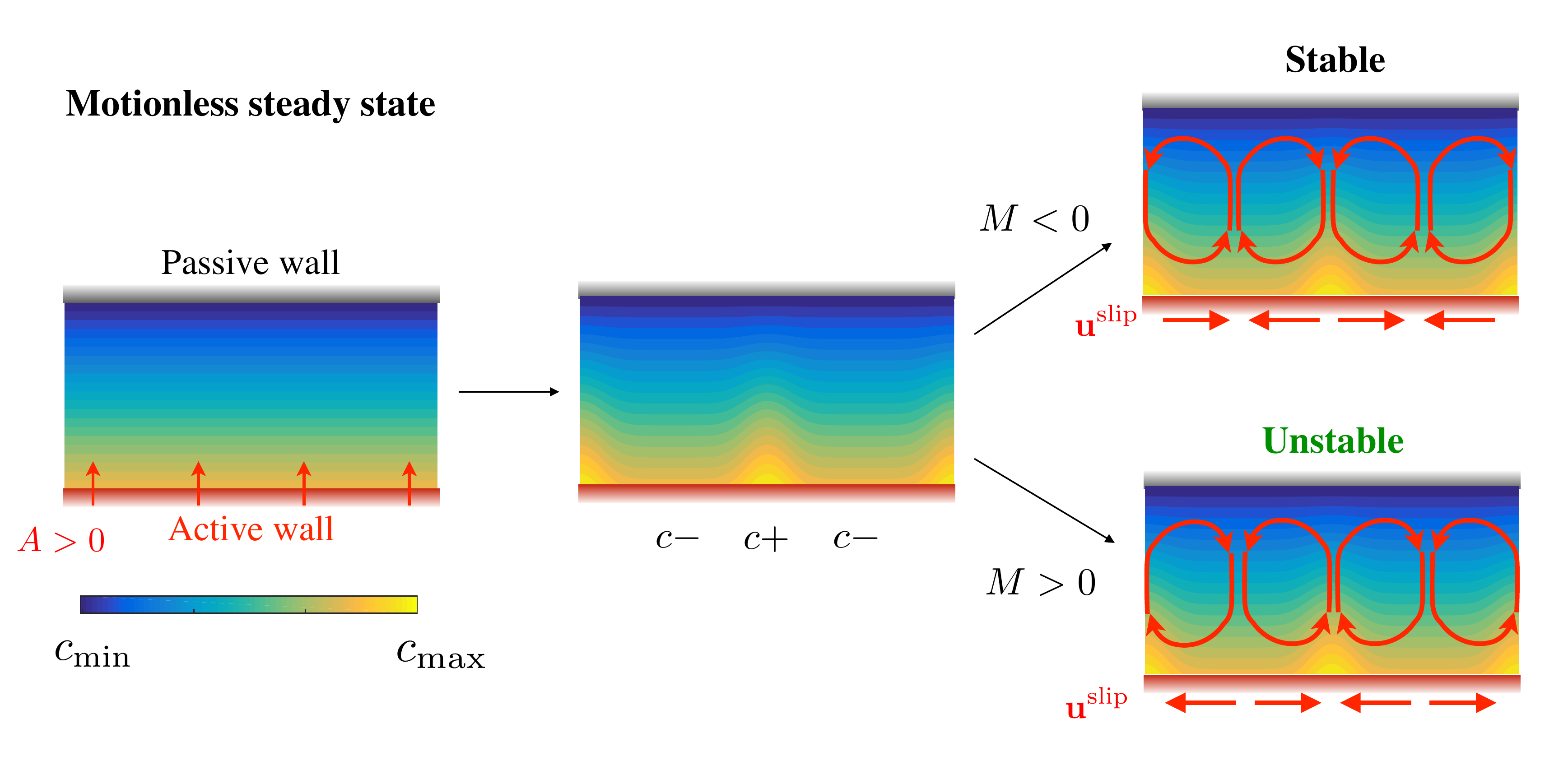}
\caption{Phoretic instability and convection in a uniform phoretic channel. Left:  Motionless steady state with no phoretic flow. Center: Perturbed initial state, with an upward plume of solute-rich fluid. Right:  Phoretic flows at the active wall can either decrease the initial perturbations (case with mobility $M<0$, top) or exacerbate  them, leading to an instability and self-sustained convective flow (case with mobility $M>0$, bottom).} \label{fig:principe}
\end{center}
\end{figure*}

However, when the solute diffuses slowly, a small perturbation of the concentration at the wall will generate a net slip flow either away or toward a region of excess solute content. In the latter case (so-called positive phoretic mobility), the resulting convective flow is expected to transport and accumulate more solute in that region, resulting in a self-sustained flow within the channel (Figure~\ref{fig:principe}). This mechanism is akin to the classical B\'enard-Marangoni convection in a thin-film~\cite{benguria1989linear,colinet2001,cloot1984nonlinear}, which is driven by a temperature difference between the two opposite surfaces and for which perturbations in the temperature distribution at the free surface generates a Marangoni flow which can drive a net convection.

Similarly to the classical B\'enard-Marangoni convection, we  demonstrate in this paper  that   phoretic phenomena can lead to a spontaneous local symmetry-breaking of the solute distribution and the creation of a self-sustained convective flow in the channel. The flow does not   pump a net amount of  fluid   in the streamwise direction but it does significantly impact the distribution of solute across the channel and its transport.  Using analytical calculations and numerical computations we  characterize the channel phoretic  instability and its long-time  saturated regimes, drawing a clear physical parallel to  B\'enard-Marangoni convection in thin films.

The paper is organized as follows. Section~\ref{sec:model} describes the simplified model considered here for the active micro-channel. The linear stability of the  steady state is then analyzed in Section~\ref{sec:linstab}, and the resulting saturated regime and its properties are characterized in Section~\ref{sec:nonlin}. Our findings are finally summarized in Section~\ref{sec:conclusions}, where we also offer some further perspectives.

\section{Model}\label{sec:model}
\subsection{Problem formulation}
We investigate here the spontaneous emergence of phoretic flows within an infinite two-dimensional channel of depth $\cal H$ with walls of homogenous chemical properties, namely (i) a chemically passive upper wall ($y=\cal H$) that also maintains a uniform solute concentration $\mathcal{C}_0$, and (ii) an active bottom wall ($y=0$) with homogeneous chemical activity $\cal A$ and phoretic mobility $\cal M$. We use  $\cal D$ to denote the diffusivity of the solute whose concentration is denoted by ${\cal C}(\rb)$ throughout the channel. The two chemical properties of the bottom wall translate into boundary conditions for the chemical concentration and flow field, namely  a fixed diffusive flux per unit area \change{${\cal A}=-\mathcal{D}\nb\cdot\grad {\cal C}$  and a phoretic slip $\ub_s={\cal M}(\mathbf{I}-\nb\nb)\cdot\grad {\cal C}$} with $\nb$ the unit  vector normal to the surface and pointing into the fluid phase. The flow is assumed to be dominated by viscous effects so the fluid velocity satisfies the incompressible Stokes equations. \change{Rewriting this two-dimensional incompressible flow in terms of a single scalar streamfunction $\psi$, $\ub=(\partial\psi/\partial y)\eb_x-(\partial\psi/\partial x)\eb_y$, $\psi$ must therefore satisfy the biharmonic equation, $\nabla^2(\nabla^2\psi)=0$~\cite{leal2007}. }

Scaling lengths by ${\cal H}$, relative concentration $c=\mathcal{C}-\mathcal{C}_0$ by $|\mathcal{A}|\mathcal{H}/{\mathcal{D}}$, fluid velocities by $\mathcal{V}=|\mathcal{AM}|/\mathcal{D}$ and times by $\mathcal{H}/\mathcal{V}$, \change{the equations for the flow field (biharmonic equation for the dimensionless  streamfunction $\psi$) and solute transport (the advection-diffusion equation for the dimensionless relative concentration $c$)} are given by
\begin{align}
\nabla^2(\nabla^2 \psi)=0,&\label{eq:Stokes}\\
\Pe\left(\pard{c}{t}+\pard{\psi}{y}\pard{c}{x}-\pard{\psi}{x}\pard{c}{y}\right)&=\nabla^2c\label{eq:advdiff},
\end{align}
with boundary conditions
\begin{align}
c=0,\qquad\pard{\psi}{y}=\psi=0\qquad \qquad&\textrm{for   } y=1,\label{eq:bc1}\\
\pard{c}{y}=-A,\quad\psi=Q_0,\quad \pard{\psi}{y}=M\pard{c}{x}\qquad&\textrm{for   } y=0,\label{eq:bc2}
\end{align}
and where we have introduced the P\'eclet number  for the solute transport, $\Pe=|\mathcal{AM}|\mathcal{H}/\mathcal{D}^2$. Here $A=\pm 1$ and $M=\pm 1$ are the dimensionless activity and mobility of the active bottom walls.

The constant $Q_0$ is the net volume flux through the channel,   a dimensionless  constant.  In the following, we assume that the problem is periodic in the streamwise direction with period $L$. \change{Using the reciprocal theorem for Stokes flows, this total volume flux $Q_0$ can in fact be obtained directly in terms of the phoretic slip at the wall $\ub_s$ as \cite{michelin2015c}
\begin{equation}
Q=\frac{1}{L}\int_{\partial \Omega}\ub_s\cdot\mathbf{f}^*\dd x,
\end{equation}
where $\mathbf{f}^*$ is the auxiliary traction force corresponding to a Poiseuille flow forced by a unit pressure gradient in the same channel geometry with no-slip boundary conditions, and the integration is performed on all the side walls of the periodic channels. Here $\ub_s$ is strictly zero at the top wall and at the bottom wall, $\mathbf{f}^*=\eb_x/2$ and $\ub_s=M\displaystyle\pard{c}{x}(x,0)\eb_x$, and  therefore, since $c$ is periodic in $x$, we obtain
\begin{equation}
Q_0=\frac{M}{2L}\int_0^L\pard{c}{x}(x,0)\dd x=0.
\end{equation}
For uniform mobility, it is therefore not possible to drive any net flow along the channel, regardless of the activity of its walls.}

\subsection{Solution to the hydrodynamic problem}
While Eqs.~\eqref{eq:Stokes}--\eqref{eq:bc2} form a non-linear system for the coupled dynamics of $c$ and  $\psi$, the streamfunction itself is a linear and instantaneous function of $c$ since inertia is negligible, 
i.e.~$\psi=\mathscr{L}[c]$ with $\mathscr{L}$ a linear and instantaneous operator. For a given distribution of concentration at the active boundary, the problem for $\psi$ can be solved for analytically in Fourier space. 
Specifically, denoting by $\hat{f}(k,y)=\int_{-\infty}^\infty f(x,y)\ee^{-\ci k x}\dd x$ the Fourier transform in $x$ of any field $f(x,y)$ in real space, we see that  $\hat\psi$ and $\hat c$ follow
\begin{align}
\left(\pard{^2}{y^2}-k^2\right)^2&\hat{\psi}=0,\\
\hat \psi(k,0)=0,\quad \pard{\hat\psi}{y}(k,0)&=\ci kM\hat{c}(k,0),\\
\pard{\hat\psi}{y}(k,1)=\hat\psi(k,&1)=0,
\end{align}
whose unique solution is given by
\begin{align}
\hat\psi(k,y)&=\ci kM\Psi(k,y)\times\hat{c}(k,0),\label{eq:stokessol}\\
\Psi(k,y)&=\frac{k(y-1)\sinh ky-y\sinh k\,\sinh [k(y-1)]}{\sinh^2 k-k^2}.
\end{align}
This means that we can write formally the streamfunction as a function of the solution concentration as a convolution
\begin{align}
\psi(x,y)&=M\int_{-\infty}^\infty K(x-x',y)c(x',0)\dd x',\\ K(u,y)&=\frac{1}{2\pi}\int_{-\infty}^\infty\ci k\Psi(k,y)\ee^{\ci k u}\dd  k.
\end{align}

\section{Linear stability analysis of the   steady state}\label{sec:linstab}
The system in Eqs.~\eqref{eq:Stokes}--\eqref{eq:bc2} admits a   steady solution which corresponds to pure diffusion of the solute across the channel and no flow through the entire channel (concentration is uniform in $x$), i.e. 
\begin{equation}
\bar{c}=A(1-y),\qquad \bar\psi=0.\label{eq:basestate}
\end{equation}

\subsection{Linearized equations}
We first focus on the linear stability analysis of this system around the   steady state from Eq.~\eqref{eq:basestate}. Decomposing $c=\bar{c}+c'$ and $\psi=\psi'$, the linearized problem for $c'$ is obtained as
\begin{align}
\Pe\pard{c'}{t}&=\nabla^2c'-A\Pe\pard{\psi'}{x},\\
c'(x,1)&=\pard{c'}{y}(x,0)=0,
\end{align}
and $\psi'$ is directly obtained from $c'$ using Eq.~\eqref{eq:stokessol}. Searching   for normal modes of the form  $c'=C(k,y)\ee^{\ci kx+\sigma t}$ with growth rate $\sigma$,  the values of $\sigma$ and $C(k,y)$ satisfy an   eigenvalue problem (for given $k$) given by
\begin{align}
\change{\pard{^2C}{y^2}}-\tilde{k}^2C=&-k^2AM\Pe\times\Psi(k,y),\label{eq:lin1}\\
\change{\pard{C}{y}(k,0)}=0,\quad &\change{C(k,1)}=0,\label{eq:lin2}
\end{align}
with $\tilde{k}^2=k^2+\Pe\,\sigma$.

\subsection{Stability threshold}
We first seek for the stability threshold by looking for neutrally-stable modes with $\sigma=0$ (so that $\tilde{k}=k$).
The general solution of Eq.~\eqref{eq:lin1} is given by
\begin{align}
\change{C(k,y)}=&\frac{AM\Pe\left[\alpha\sinh[k(y-1)]+\beta\cosh ky+G(k,y)\right]}{4(\sinh^2k-k^2)},\\
G(k,y)=&ky^2\sinh k\cosh[k(y-1)]+k(y-1)\sinh[ky]\nonumber\\
&-k^2(y-1)^2\cosh[ky]-y\sinh k\sinh[k(y-1)].
\end{align}
Imposing $\change{C(k,1)}=0$ and $\displaystyle\change{\pard{C}{y}(k,0)}=0$   leads to
\begin{align}
\beta=-k\tanh k,\qquad \alpha=-\frac{\sinh^2k+k^2}{k\cosh k}.
\end{align}
\begin{figure}
\begin{center}
\includegraphics[width=.55\textwidth]{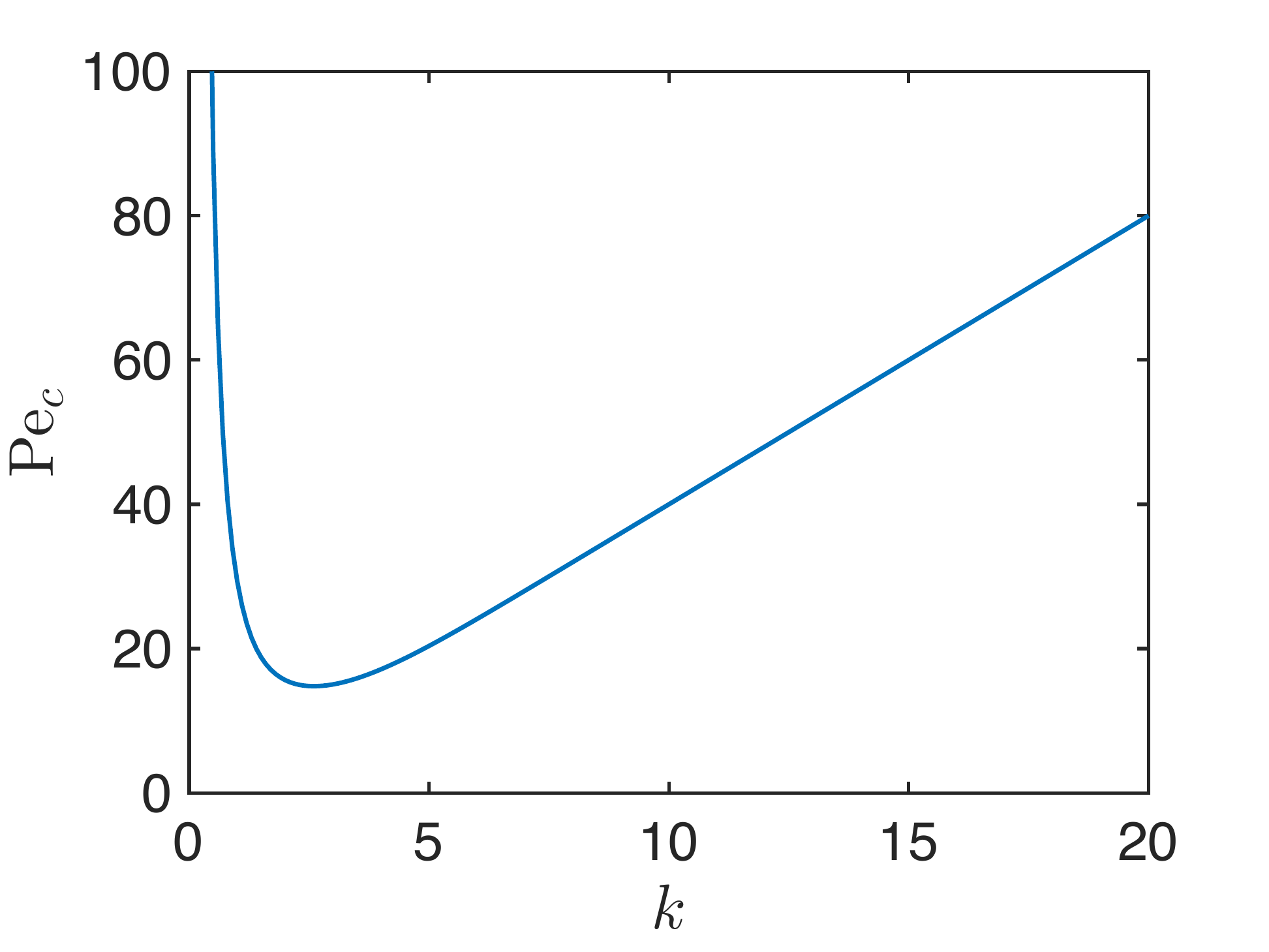}
\caption{Neutral stability curve of the channel phoretic instability. Dependence of the critical P\'eclet number, $\Pe_c$,  on the  wavenumber of the Fourier mode of the perturbation, $k$, for $AM=1$.}\label{fig:pec}
\end{center}
\end{figure}
Finally, normalising the eigenfunction such that $\change{C(k,0)}=1$ provides the values $\Pe_c(k)$ for which a neutrally-stable mode of wave number $k$ exists, namely
\begin{equation}
\Pe_c(k)=\left(\frac{4}{AM}\right)\frac{k(\sinh^2 k-k^2)}{\tanh k \sinh^2 k-k^3}\cdot
\end{equation}
When $AM=1$, $\Pe_c(k)>0$ and its dependence on $k$ is plotted in Figure~\ref{fig:pec}. Solute advection is the driving mechanism of any potential instability. Thus, when $\Pe$ is small, the solute dynamics is dominated by diffusion which induces an overdamped relaxation of any perturbation, and all modes are linearly stable. Unstable modes can only develop (i) for small-enough diffusion (i.e.~above a critical $\mbox{Pe}\geq\mbox{Pe}_c$) and (ii) only if $AM=1$ so that local concentration perturbations are reinforced by the resulting phoretic advection. When $AM=-1$, all the modes are linearly stable regardless of the P\'eclet number. In what follows, we focus exclusively on the case $AM=1$ that can potentially lead to an instability, and thus set $A=M=1$. 

As seen in Figure~\ref{fig:pec}, above a critical value of the P\'eclet number,  $\mbox{Pe}\geq \Pe_0\approx 14.8$, the equation $\Pe_c(k)=\Pe$ admits two distinct solutions $k_1<k_2$ and all the modes with $k_1<k<k_2$ are linearly unstable, while those with $k<k_1$ or $k>k_2$ are linearly stable. This finite wavelength instability is not surprising physically, and its origin is similar to the physical mechanisms underlying the classical B\'enard-Marangoni instability. Perturbation modes with high $k$, i.e.~when the channel width is much larger than the wavelength, are  damped by diffusion in the streamwise direction. In contrast,  modes with low $k$ correspond to very elongated rolls for which the channel width is too small for any significant longitudinal gradient to develop and drive a net flow.
This damping of the small- and large-$k$ modes is confirmed by the divergence of $\mbox{Pe}_c(k)$ in both limits. Indeed, we have asymptotically
\begin{align}
\textrm{for    }k\rightarrow 0,&\qquad \Pe_c(k)\sim \frac{20}{k^2},\\
\textrm{for    }k\rightarrow\infty,&\qquad \Pe_c(k)\sim 4k.
\end{align}
The minimum value,  $\Pe_0\approx 14.8$, has an associated wavenumber   ${k_0\approx 2.57}$, and is the critical P\'eclet number below which no instability can develop.

\begin{figure}[t]
\begin{center}
\begin{tabular}{c}
\includegraphics[width=.5\textwidth]{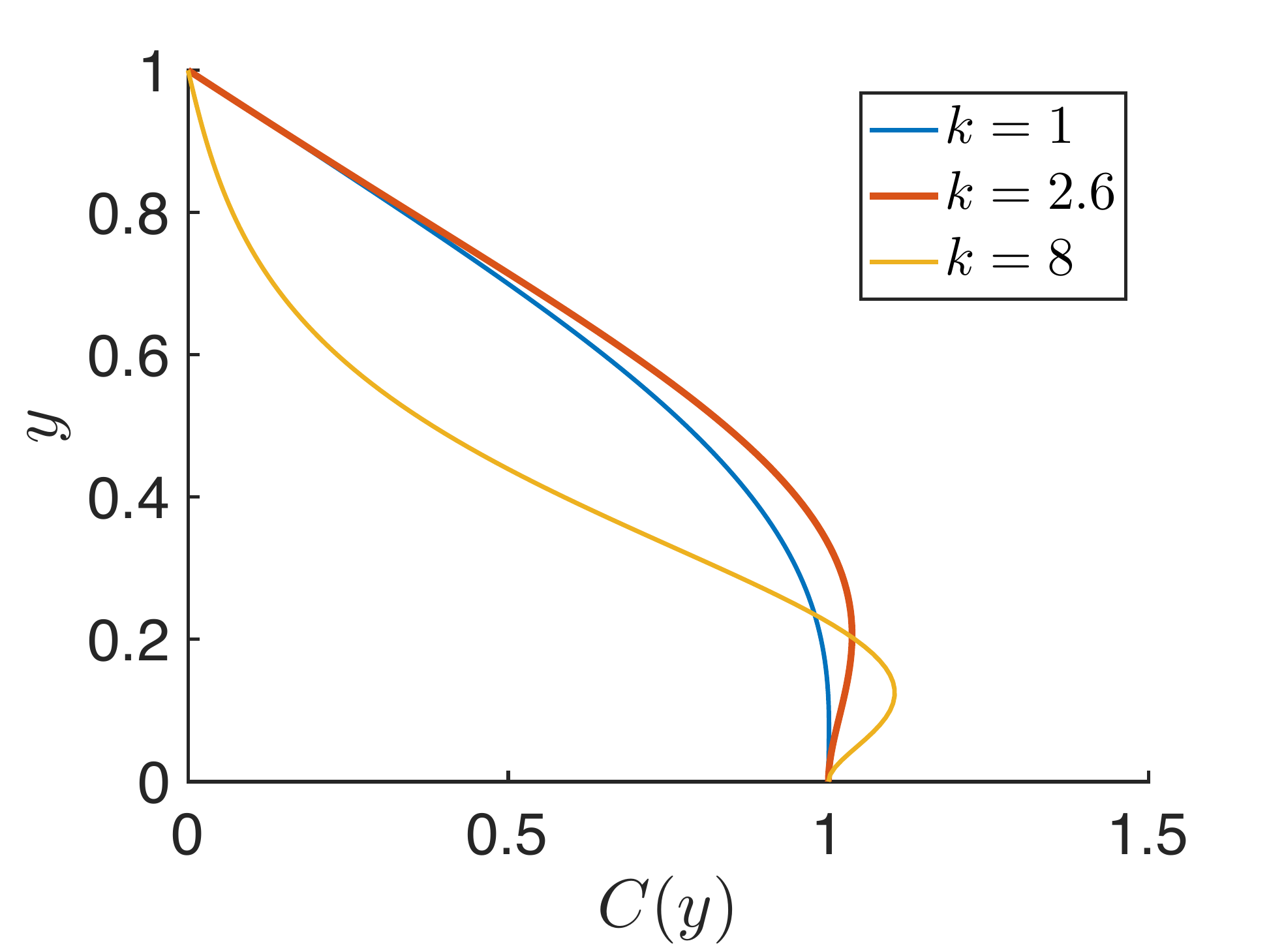}\\
\includegraphics[width=.55\textwidth]{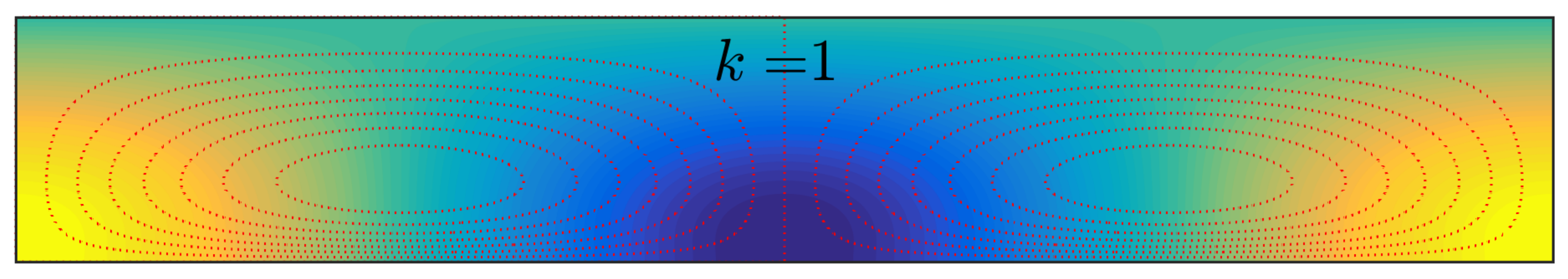}\\
\includegraphics[width=.55\textwidth]{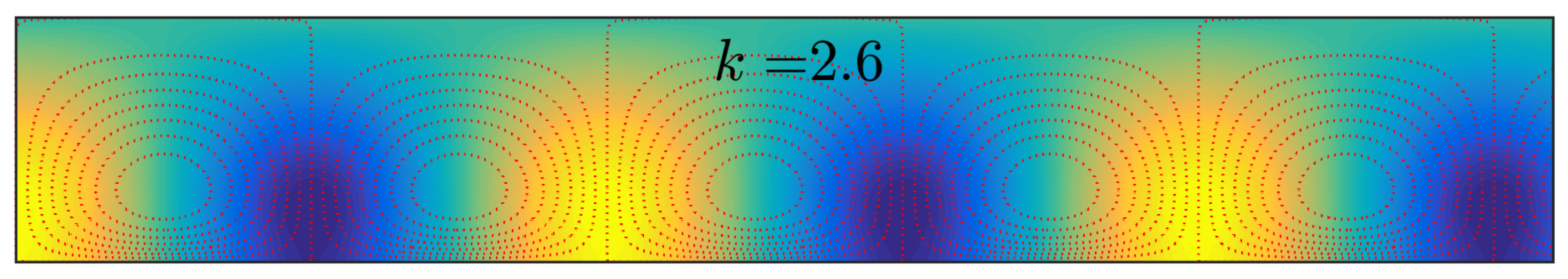}\\
\includegraphics[width=.55\textwidth]{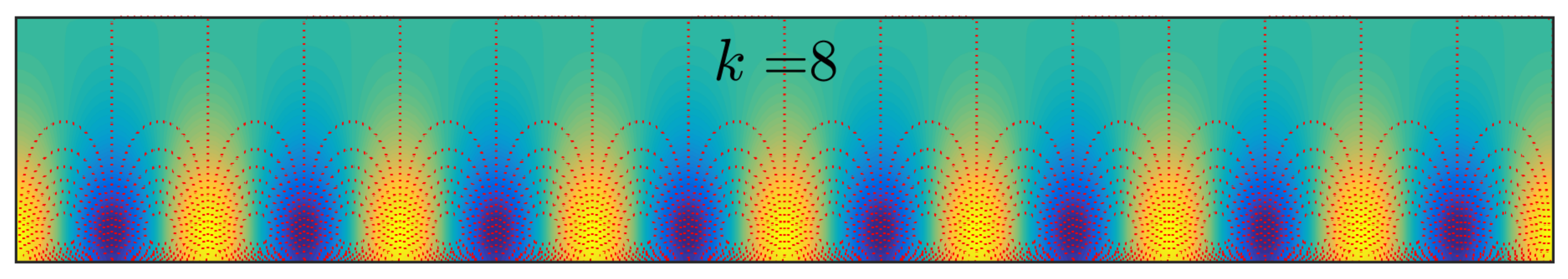}
\end{tabular}
\caption{
\change{(Top)}: Vertical structure of the neutrally-stable eigenmode for three different values of $k$. 
\change{(Bottom)}: Concentration perturbation and resulting streamlines for the neutrally-stable eigenmodes of increasing $k$. Lighter (resp. darker) color depicts a positive (resp. negative) concentration perturbation. Note that the corresponding value of the P\'eclet number is therefore different for each $k$, namely $\Pe_c=29.4$, $14.8$ and $32$ for $k=1$, $k=2.6\approx k_0$ and $k=8$ respectively. In all cases, $AM=1$.}\label{fig:eig_mode}
\end{center}
\end{figure}

The structure of the corresponding neutrally-stable eigenmode shows the emergence of regions of increased concentration along the active bottom wall, leading to a net phoretic slip (i) toward the regions of higher concentration (light color) where the flow is oriented upward  and (ii) away from regions of reduced concentration (dark color) where the flow is oriented downward (Figure~\ref{fig:eig_mode}). The vertical structure of the eigenmode  depends on its wavenumber, $k$, and hence the aspect ratio of the counter-rotating flow cells. For long waves ($k\lesssim k_0$), the concentration perturbation varies monotonously across the width of the channel while for shorter waves ($k\gtrsim k_0$) the maximum perturbation amplitude is reached at a finite distance from the wall.

\subsection{Growth rate of unstable modes}
Away from the critical P\'eclet number, it is important to consider the general eigenvalue problem in Eqs.~\eqref{eq:lin1}--\eqref{eq:lin2} and its solution for $k\neq \tilde k$ (i.e.~$\sigma\neq 0$).
The generic solution for $C$ is given by 
\begin{align}
\change{C(k,y)}=&\frac{k^2AM\Pe\left[\alpha\sinh[\tilde{k}(y-1)]+\beta\cosh\tilde{k}y+\tilde{G}(k,y)\right]}{(\sinh^2k-k^2)},
\end{align}
with
\begin{align}
\tilde{G}(k,y)=&\frac{2k^2\cosh(ky)-2k\sinh k\cosh[k(y-1)]}{(\tilde{k}^2-k^2)^2}\nonumber\\
&+\frac{k(y-1)\sinh(ky)-y\sinh k\sinh[k(y-1)]}{(\tilde{k}^2-k^2)}\cdot
\end{align}
Note that the solution above is valid regardless of the sign of $\tilde k^2$ and when $\tilde k^2 < 0$, $\tilde k$ is imaginary and one finds the solution using $\sinh (\ci z)=i\sin z$ and $\cosh (\ci z)=\cos \ci z$. 

The conditions $\change{C(k,1)}=0$ and $\change{\displaystyle\pard{C}{y}(k,0)}=0$ impose respectively
\begin{align}
\beta&=-\frac{\tilde{G}(k,1)}{\cosh\tilde{k}},\qquad
\alpha=-\frac{1}{\tilde{k}\cosh\tilde{k}}\pard{\tilde{G}}{y}(k,0),
\end{align}
and finally $\change{C(k,0)}=1$ provides an equation for $\tilde{k}$ (and $\sigma$) as a function of $k$ and $\Pe$ which is solved numerically.

\begin{figure}
\begin{center}
\includegraphics[width=.55\textwidth]{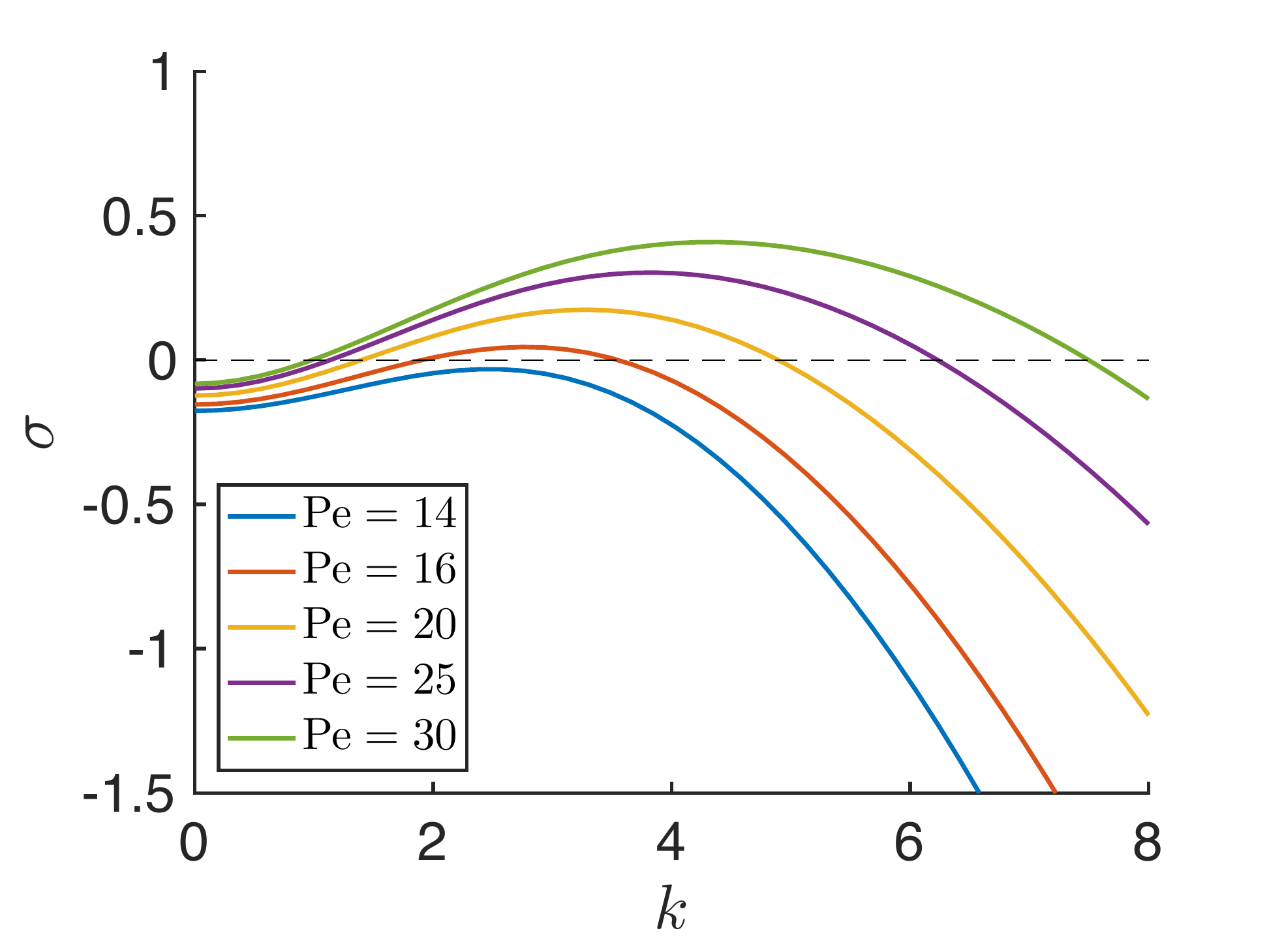}
\caption{Unstable modes of channel phoretic flow. 
Growth rate, $\sigma$, of a Fourier mode with wavenumber, $k$, for increasing values of P\'eclet number, $\Pe$, in the case $AM=1$.}\label{fig:sigma}
\end{center}
\end{figure}

\begin{figure}[t]
\begin{center}
\includegraphics[width=.55\textwidth]{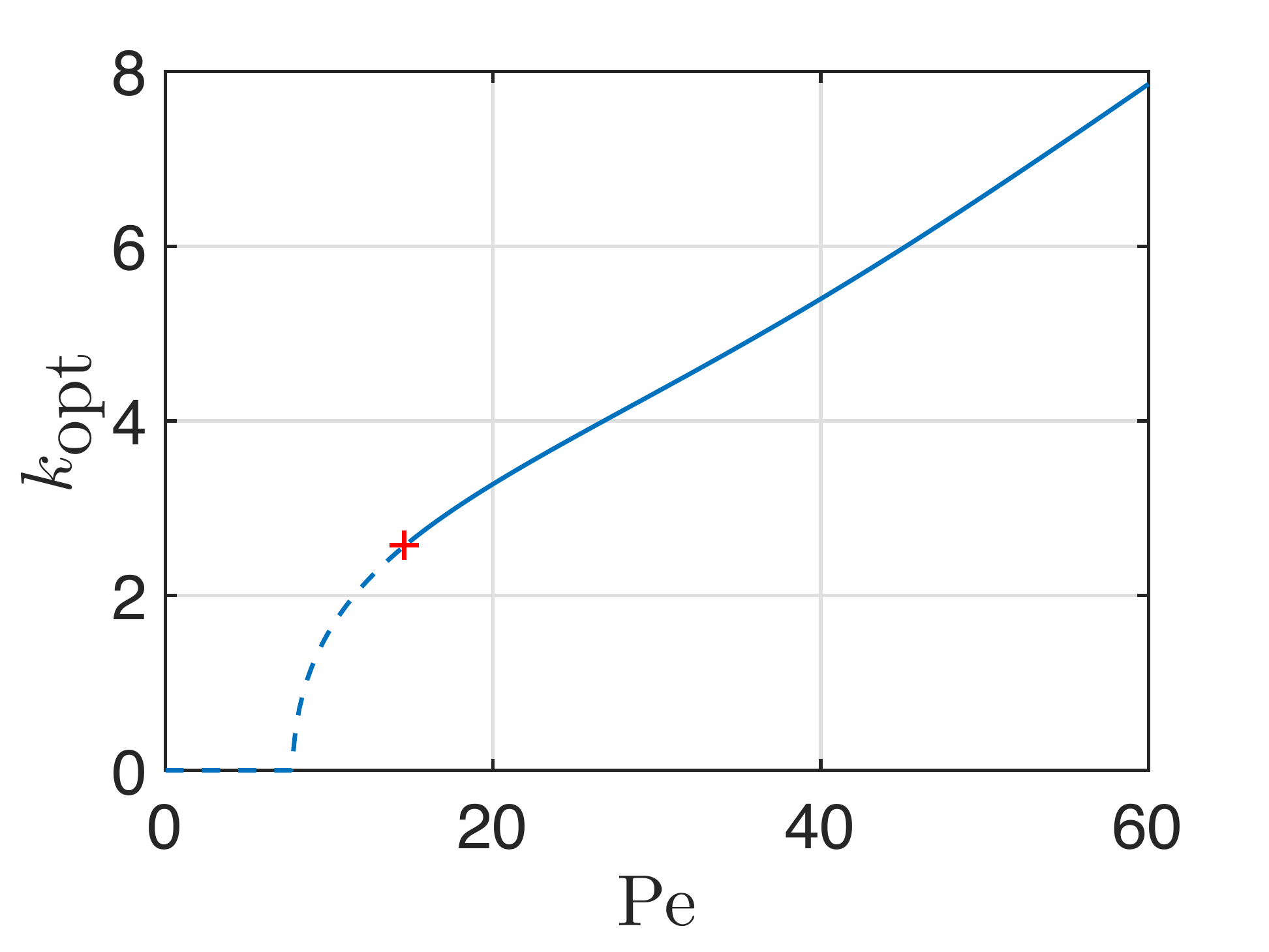}
\caption{Dependence of the wavenumber of the most unstable mode,  $k_\textrm{opt}$,  i.e.~the one with the largest growth rate, with the P\'eclet number, $\Pe$, in the case  $AM=1$. The line is solid when this mode is unstable and dashed when it is stable (the red cross corresponds to the bifurcation point ($k_0,\Pe_0$).}\label{fig:kopt}
\end{center}
\end{figure}

The dependence of the growth rate, $\sigma$, on the wavenumber, $k$, is shown for increasing values of the P\'eclet number, $\Pe$, in Figure~\ref{fig:sigma}. These results confirm the existence of a minimum value $\Pe_0\approx 14.8$, below which all modes are stable. For $\Pe=\Pe_0$, a single wave number is neutrally stable ($k_0\approx 2.57$). For $\Pe\geq \Pe_0$, an increasing range of unstable wavenumbers is found, and the most unstable wavenumber $k_\textrm{opt}$ increases with $\Pe$ (see Figure~\ref{fig:kopt}).

\subsection{Stability analysis in a periodic channel}
\begin{figure}[t]
\begin{center}
\begin{tabular}{cc}
\includegraphics[width=.45\textwidth]{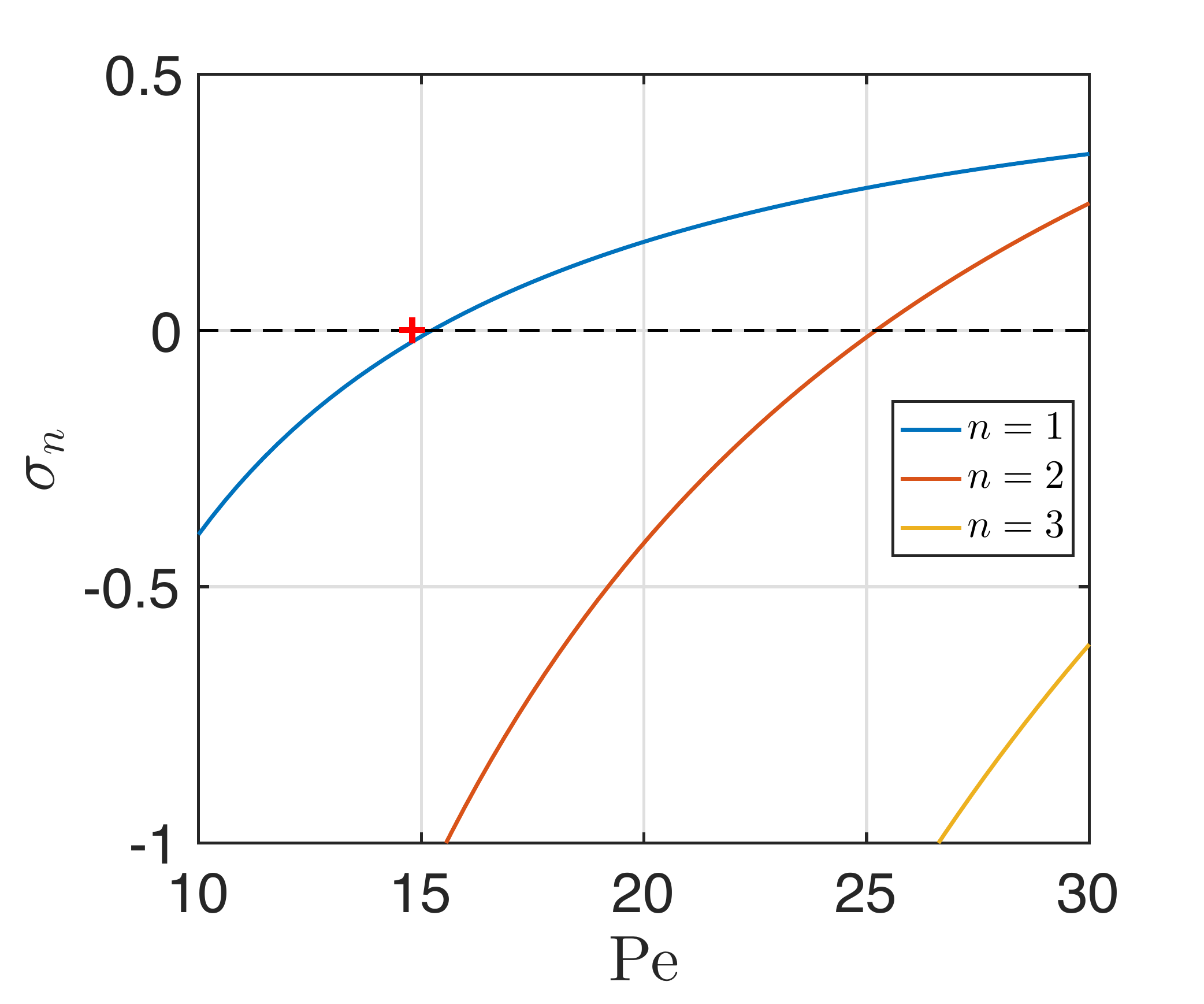} &
\includegraphics[width=.45\textwidth]{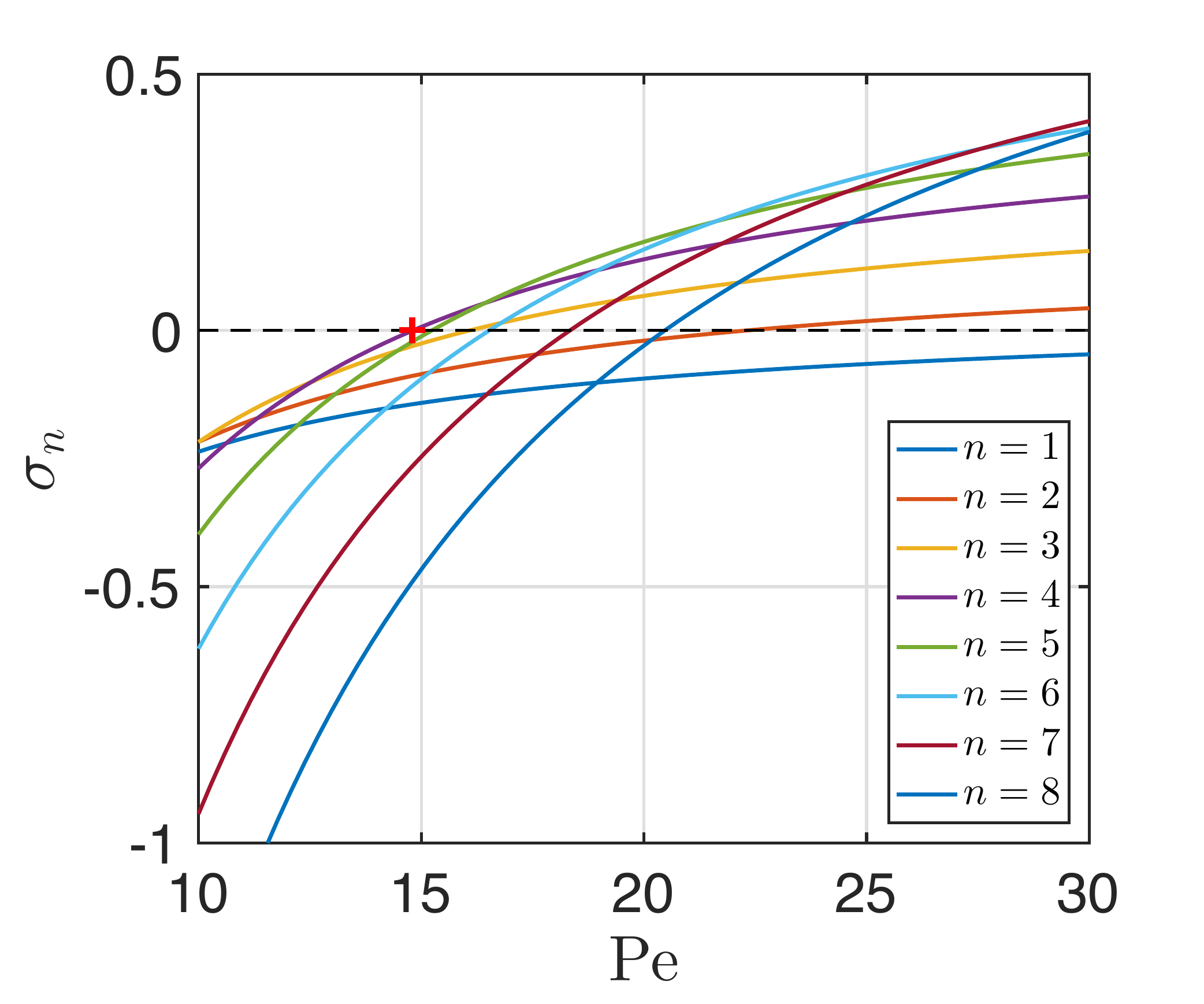}
\end{tabular}
\caption{Unstable modes of  phoretic flow for a periodic channel. 
Dependence of the growth rate of the mode of order $n$ (i.e.~with $n$ periods) in a channel of longitudinal period $L=2$ (left) and $L=10$ (right), in the case $AM=1$. The red cross indicates the position of the instability threshold in an infinite domain ($\Pe_0\approx 14.8$).}\label{fig:growth_L10}
\end{center}
\end{figure}

When a periodic channel is considered (with lengthwise period $L$), only a discrete set of modes can be found, namely $k_j=2\pi j/L$ with $j\in \mathbb{N}$. For a fixed length 
(e.g.~$L=10$), the most unstable wave number will vary as $k\propto\Pe$ as the P\'eclet number 
 is increased (Figure~\ref{fig:growth_L10}). For small $L$, the value of the instability threshold. i.e.~the minimum $\Pe$ beyond which at least one mode is unstable, can be significantly different from $\Pe_0$ (infinite domain), while for sufficiently large   $L$ it is well reproduced.

\section{Nonlinear dynamics in a periodic channel}\label{sec:nonlin}
The results of the previous section identified a critical P\'eclet number beyond which the   steady state corresponding to a uniform distribution of the concentration along the channel wall becomes unstable, following a mechanism similar to  the classical B\'enard-Marangoni instability. For $\Pe\geq \Pe_0$, when a small perturbation is introduced to the steady state solution, a finite range of eigenmodes with finite $k$ and positive growth rates are expected to grow exponentially, with one of the modes (that with maximum growth rate) becoming dominant over the slower-growing other modes. When perturbations to the   steady state are no longer negligible,  the nonlinear advection of the concentration perturbation by the phoretic flows is expected to set in and drive the nonlinear saturation of the instability into a steady state where the phoretic flows along the bottom wall greatly modify the structure of the concentration field. In this section, we solve the
 full nonlinear problem for the concentration and flow field, Eqs.~\eqref{eq:Stokes}--\eqref{eq:bc2},  within a periodic channel of non-dimensional length $L$ (scaled by the channel width) numerically.

\subsection{Numerical solution of the time-dependent and steady-state problems}\label{sec:methods}
Two types of numerical simulations are performed to analyze the nonlinear solute-flow dynamics: (i) a time-dependent evolution of the steady state, Eq.~\eqref{eq:basestate}, from a small perturbation, and (ii) a direct search of the steady solutions of the full problem. The methods considered in each case are outline below.

\subsubsection{Unsteady simulations}  \change{The Stokes flow problem is linear and instantaneous  and,  at each instant, the streamfunction $\psi(x,y,t)$ can   be expressed analytically in Fourier space in terms of the wall concentration, $c(x,0,t)$, using Eq.~\eqref{eq:stokessol}. The time-dependent advection-diffusion problem, Eq.~\eqref{eq:advdiff}--\eqref{eq:bc2}, can therefore be rewritten formally as a nonlinear partial differential Equation (PDE) for $c$
\begin{equation}\label{eq:pde}
\pard{c}{t}+\mathcal{N}(c)=0,
\end{equation}
where $\mathcal{N}$ is a nonlinear spatial differential operator that accounts for the convective and diffusive transport. This unsteady PDE is marched in time using second-order centered finite differences to evaluate the spatial derivatives in $\mathcal{N}$, while time-integration is handled using a fourth-order Runge-Kutta scheme. }

 For each value of the P\'eclet number considered, and unless specified otherwise, the simulation is initiated by adding a random perturbation to the   steady state \change{$\bar{c}=1-y$}, in the form $\change{c(x,y,t=0)}=\bar{c}(y)+\varepsilon \xi(x)(1-y^2)$, with $\xi(x)$ an $O(1)$ random perturbation.

\subsubsection{Steady simulations} \label{sec:methods_simon}

In addition to the initial value time-dependent computations described above,
we have also implemented a direct solver that searches for non-trivial steady
states of Eqs.~\eqref{eq:Stokes}--\eqref{eq:bc2} after dropping the $\partial c/\partial t$ term from Eq.~\eqref{eq:advdiff}, \change{i.e.~that identifies the solutions $\tilde{c}$ of $\mathcal{N}(\tilde{c})=0$ in Eq.~\eqref{eq:pde}. }
This is especially useful given the co-existence of distinct
cellular states at the same parameter values that were reported in Figures 7 and 8.
Spectral methods were used and Fourier-Chebyshev collocation methods were 
constructed and implemented \change{in order to compute the different spatial derivatives involved in $\mathcal{N}$ in the $x$ and $y$ directions, respectively}. Spatial periodicity renders such treatments 
spectrally accurate in both the $x$ and $y$ discretizations, and the resulting
algorithms are highly accurate and efficient. Fourier and Chebyshev differentiation
matrices were used to produce a large system of nonlinear equations for the
unknown stream function and concentration fields at the collocation points.
(As mentioned earlier, nonlinearity arises due to convective coupling at non-zero $Pe$.)
The resulting system was solved using a Newton-Raphson iteration, coupled with 
a continuation method to construct bifurcation diagrams of non-trivial solutions
as $Pe$ varies such as that given in Fig. 9.

 Our computational search generically resulted 
in the coexistence of multiple states at the same Peclet number. Not all of these are
stable, however, and so the stability of all computed branches was also \change{determined by linearizing the time-dependent problem Eq.~\eqref{eq:pde} around each identified steady solution $\tilde{c}$, as follows
\begin{equation}\label{eq:linstab}
\pard{c'}{t}+\mathcal{J}(\tilde{c})c'=0
\end{equation}
with $c'=c-\tilde{c}$ a small perturbation and $\mathcal{J}(\tilde{c})$ the Jacobian of $\mathcal{N}$ evaluated at $\tilde{c}$. Analogous discretization methods were used and the stability equation, Eq.~\eqref{eq:linstab}, was thus recast into a
computational generalized eigenvalue problem.} This enables us to classify stable and unstable
states and in particular to investigate whether different states emerging at the same Peclet number
can be stable simultaneously as discussed in the results of Figures 7 and 8.

\begin{figure*}[h!]
\begin{center}
\includegraphics[width=.97\textwidth]{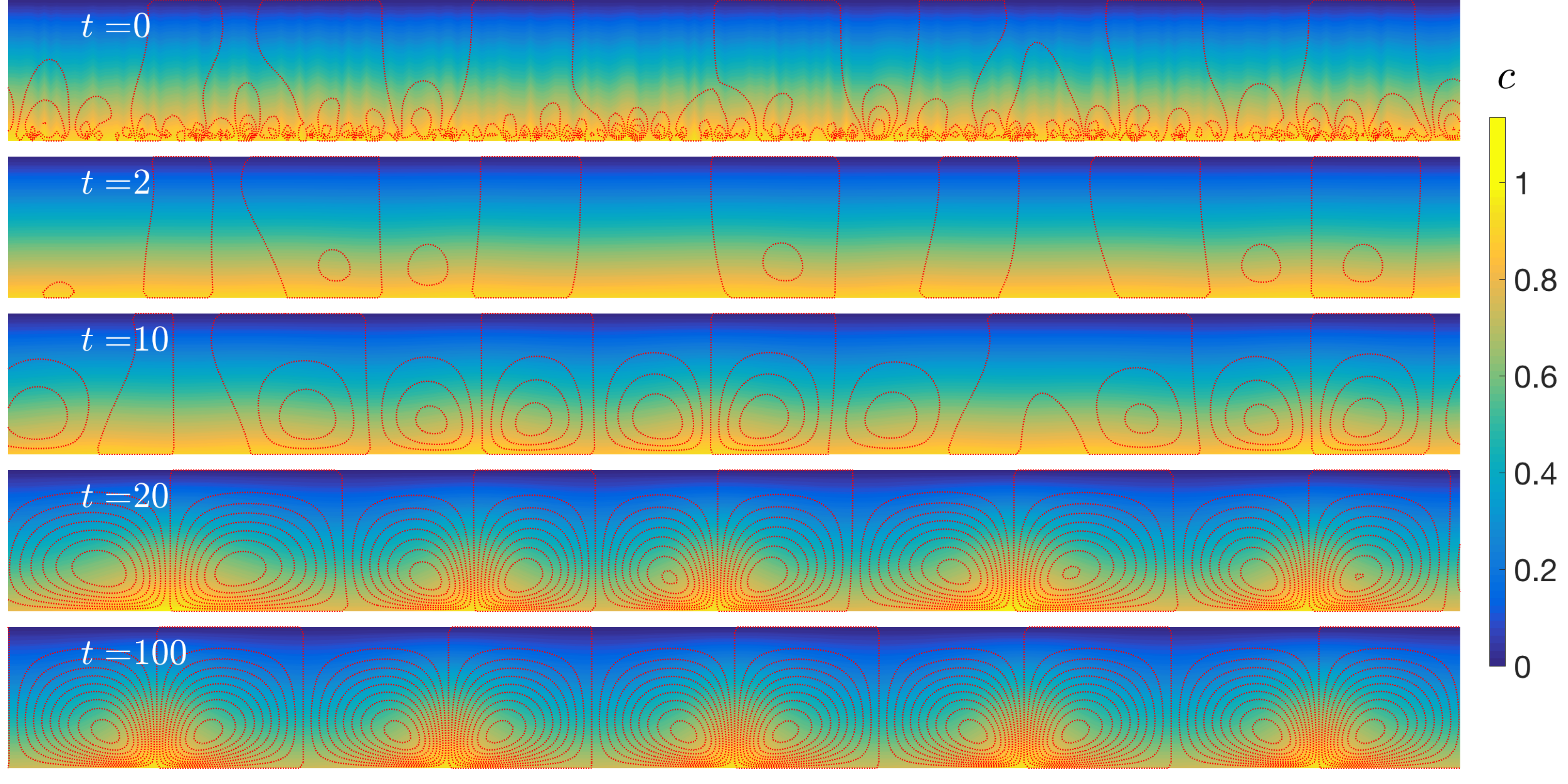}
\caption{Evolution of the concentration (color) and streamlines (dotted lines) in a periodic uniform phoretic channel with $AM=1$, $\Pe=20$ and $L=10$. The initial condition is a small random perturbation of the   steady state $\bar{c}=1-y$.}\label{fig:L10Pe20}
\end{center}
\end{figure*}

\subsection{Phoretic convection in a uniform periodic channel}

We show in Figure~\ref{fig:L10Pe20} the time-evolution  of the concentration and flow fields within the channel above the critical P\'eclet number. Initially, a finite random perturbation is added to the   steady state, leading to a complex, weak flow pattern near the active wall. 

Starting from the perturbed   steady state with no fluid motion, the first phase of evolution is characterized by (i) the rapid diffusion-driven damping of the shortest wavelengths of the perturbation and (ii) the accumulation of solute  in the convergence regions of the slip flow along the active boundary. This results in the emergence of counter-rotating cells, driven by the phoretic forcing at the bottom boundary, with $O(1)$ aspect ratio,i.e.~their longitudinal wavelength approximately scales with the channel width. Each flow cell appears to be limited in the streamwise direction by a local maximum and  minimum of the surface concentration along the active wall and is driven by the unidirectional phoretic slip joining them, which in turn sets the bulk flow into motion. Near maxima in solute concentration, the convergence of the phoretic forcing leads to an upward flow that drives fluid with higher solute content into the bulk region, while near local concentration minima the divergence of the phoretic forcing results in a downward flow that reduces the $y$-averaged concentration across the channel at that location.  In a second (much slower) phase, these cells  interact by the convective flows they generate until a steady state is reached with a finite number of cells, a final state that is quite robust with respect to the initial perturbation.

\begin{figure*}[t]
\begin{center}
\includegraphics[width=.97\textwidth]{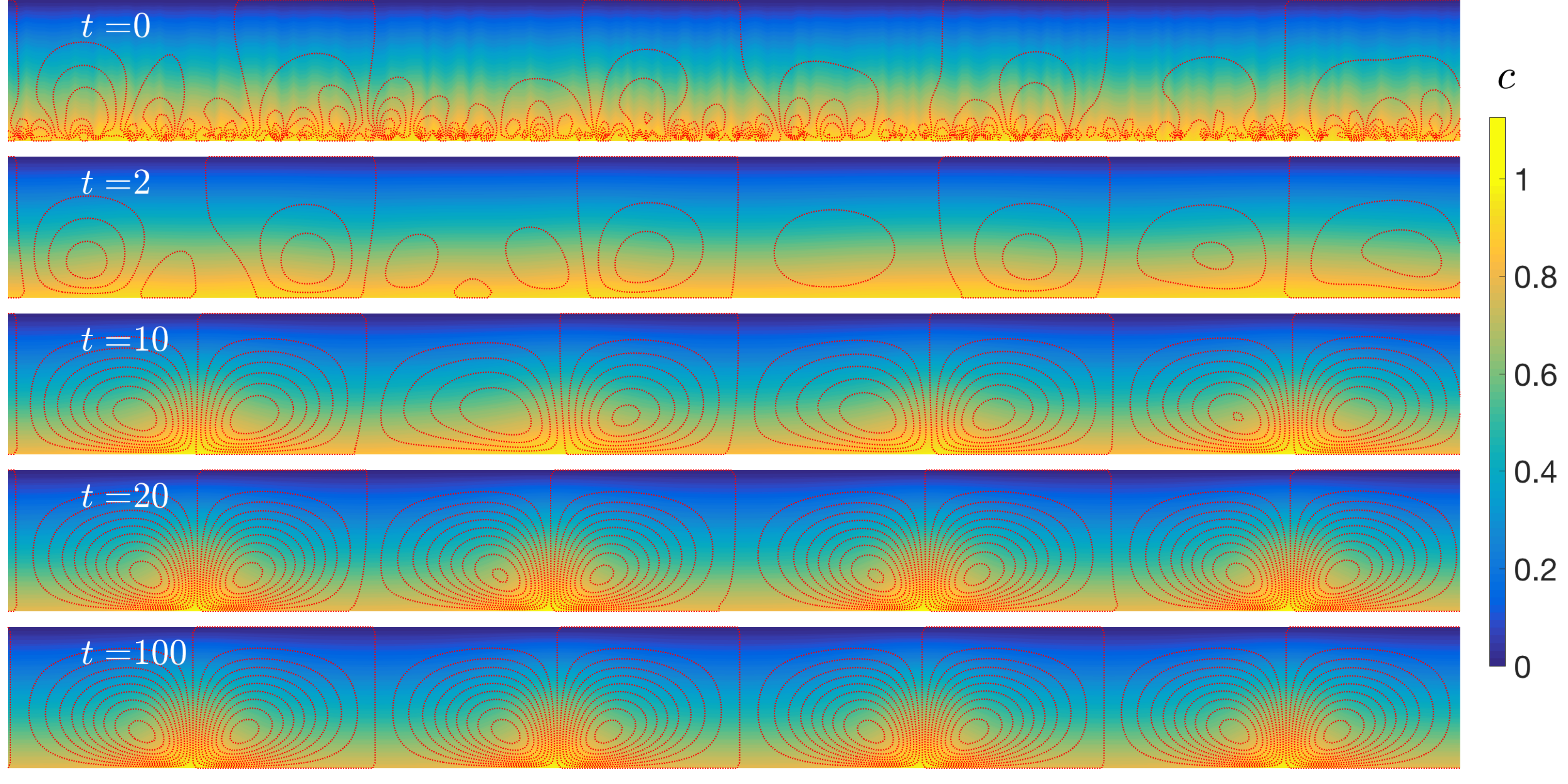}
\caption{Same as Figure~\ref{fig:L10Pe20} with a different initial random perturbation to the steady state $\bar{c}=1-y$.}\label{fig:L10Pe20_2}
\end{center}
\end{figure*}

In Figure~\ref{fig:L10Pe20}, the length of the channel (i.e.~its imposed periodicity) was set to $L=10$ and $\Pe=20$. For this combination of parameters, the linear stability analysis predicts that modes with $n=5$ cells have the largest growth rate and are therefore expected to dominate the dynamics at least initially. This is indeed observed here as the final state includes an integer number of roughly similar concentration and flow patterns  with $n=5$.  Importantly, this selection of the final nonlinear pattern overlooks the complexity imposed on such a system by the final longitudinal size of the domain, a feature that is well-known to be a generic property of Rayleigh-B\'enard-Marangoni-type flows. Indeed, as shown on Figure~\ref{fig:L10Pe20_2}, when  we consider the same domain with a slightly different initial perturbation of the steady state, we obtain   a final steady state that is fundamentally different in structure (here 4 pairs of rotating cells when 5 pairs were obtained in Figure~\ref{fig:L10Pe20}). This results is not simply a transient regime, before the system would converge again to five pairs of cells. Instead, it suggests the existence of multiple steady solutions, a result that is  confirmed in the next section. The selection by the system of one of those steady solutions depends sensitively  on the nonlinear interactions between  the different modes in the saturation process. 

\subsection{Steady state flows and modified solute transport}
\begin{figure}[t]
\begin{center}
\includegraphics[width=.6\textwidth]{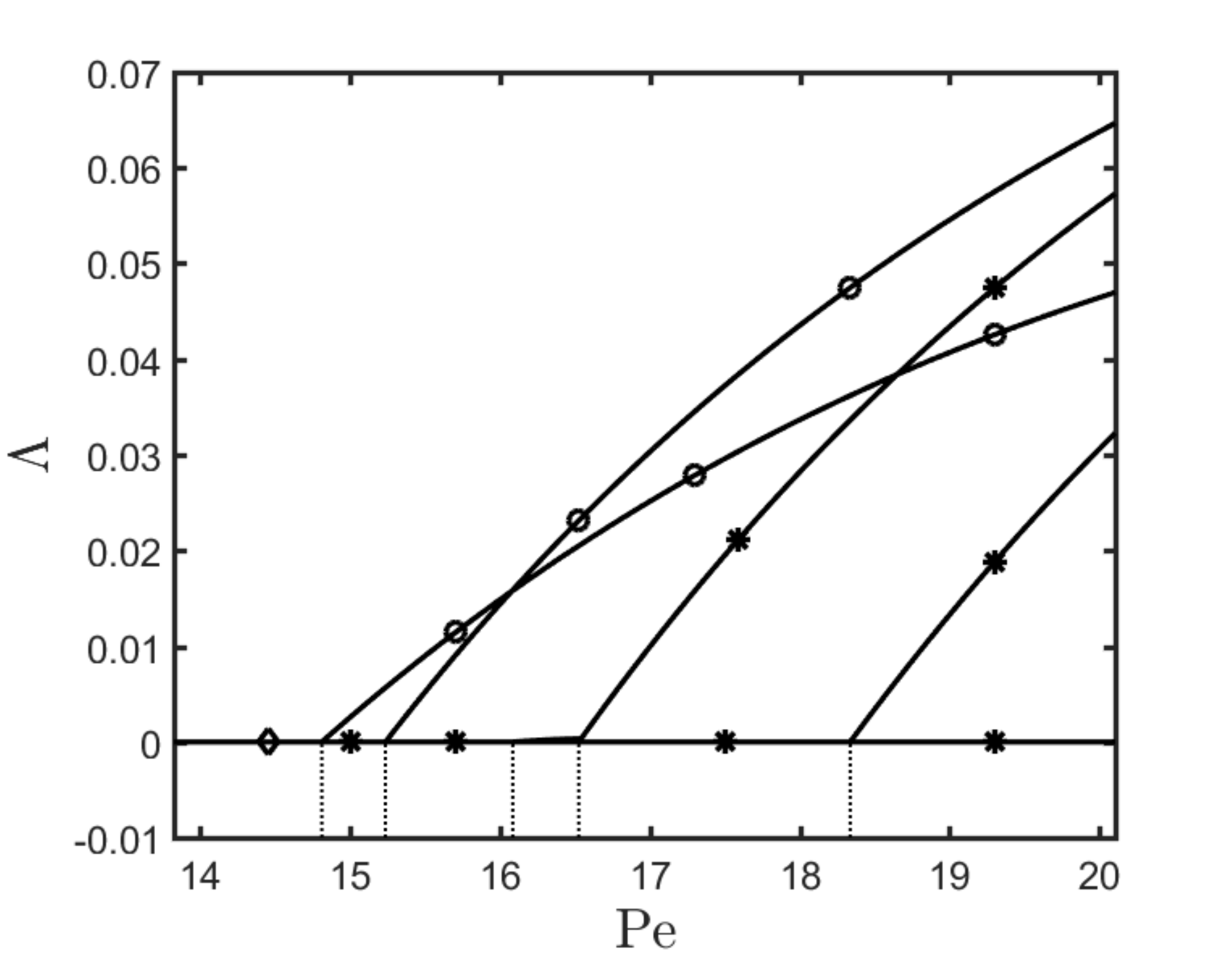}
\caption{Dependence of  wall flow magnitude,  $\Lambda$, with P\'eclet number, $\Pe$, in the   steady-state solutions for $AM=1$ and $L=10$ as obtained from the steady-state search described in Section~\ref{sec:methods_simon}. The dashed line segments correspond to the values of $\Pe$ at which the bifurcation is predicted by the linear stability analysis in a channel of finite size ratio (see Figure~\ref{fig:growth_L10}). The open circles denote a linearly stable branch, while stars denote an unstable branch.}\label{fig:bifurcation}
\end{center}
\end{figure}

To analyze this question further, we now turn to a direct search of the steady state solutions of the system, as described in Section~\ref{sec:methods_simon}. The  bifurcation diagram obtained from our steady simulations is plotted on Figure~\ref{fig:bifurcation} where the intensity of each solution is characterized by a single measure, $\Lambda$,  defined as the mean value of the phoretic slip velocity $u_s$ along the active wall,
\begin{equation}
\Lambda\equiv\frac{1}{L}\int_0^L\left.u_s^2\right|_{y=0}\dd x.\label{eq:lambda}
\end{equation}

These calculations confirm the result of the linear stability analysis with the successive emergence of a new non-trivial solution for the critical value of $\Pe$ identified in Figure~\ref{fig:growth_L10}, and an increasing number of pairs of counter-rotating flow cells, $n$, starting with $n=4$. It also confirms the existence of multiple stable branches with $n=4$ and $n=5$, and thus the possibility to obtain different steady states depending on the particular initial conditions considered, as observed in Figures~\ref{fig:L10Pe20} and \ref{fig:L10Pe20_2}.

Yet, despite their differences, the two final steady states illustrated on those figures share some similar macroscopic characteristics. In the following, we focus on two particular measures of the effect of phoretic convection. The first one, $\Lambda$ introduced in Eq.~\eqref{eq:lambda}, is a measure of the intensity of the resulting (mixing) flow field forced by the active boundary. The second measure is the mean concentration within the entire channel, $C_m$. 

\begin{figure}[t]
\begin{center}
\begin{tabular}{cc}
\subfigure[\,Mean concentration and flow magnitude]{\includegraphics[width=.48\textwidth]{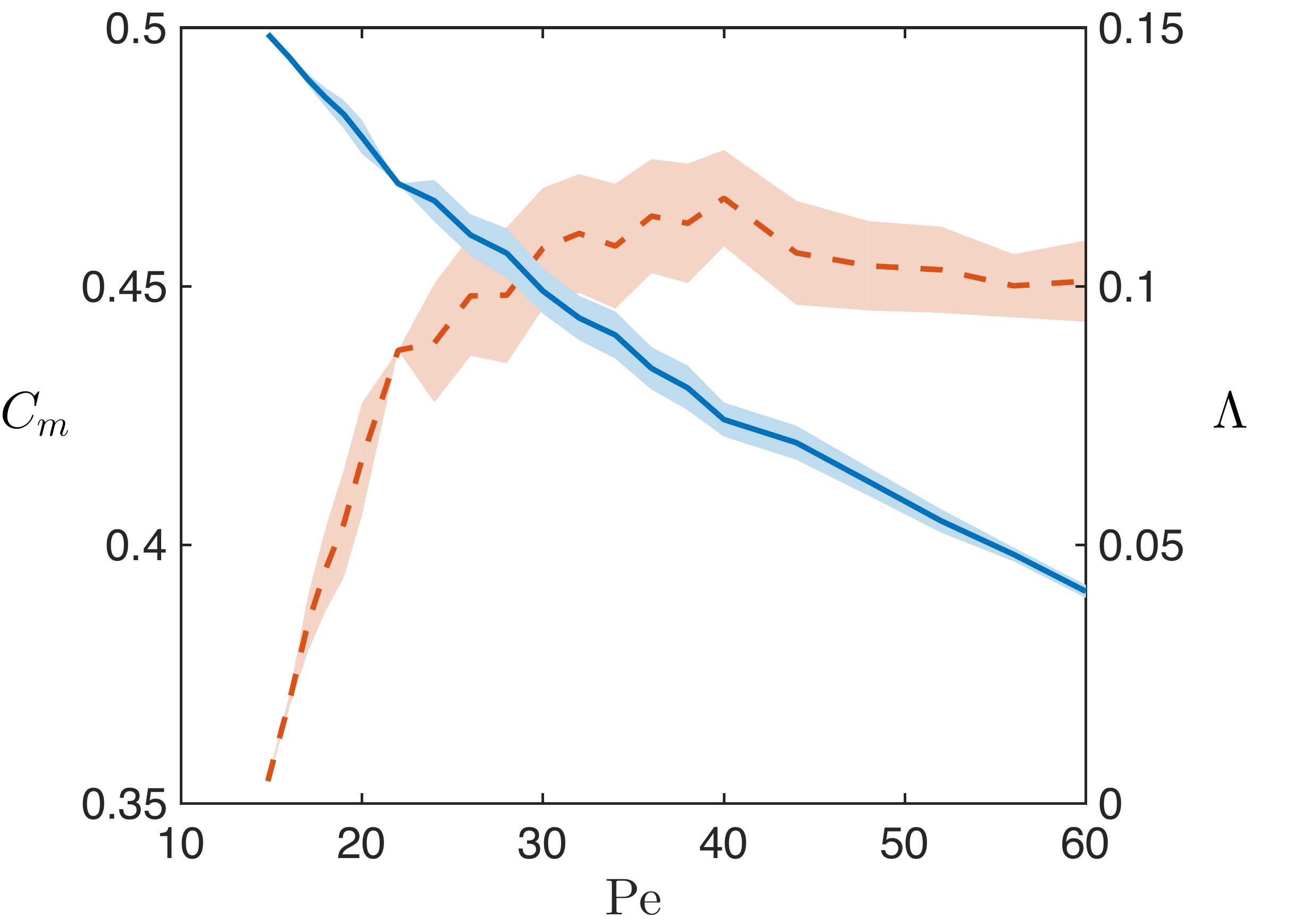}}&
\subfigure[\,Horizontally-averaged concentration]{\includegraphics[width=.48\textwidth]{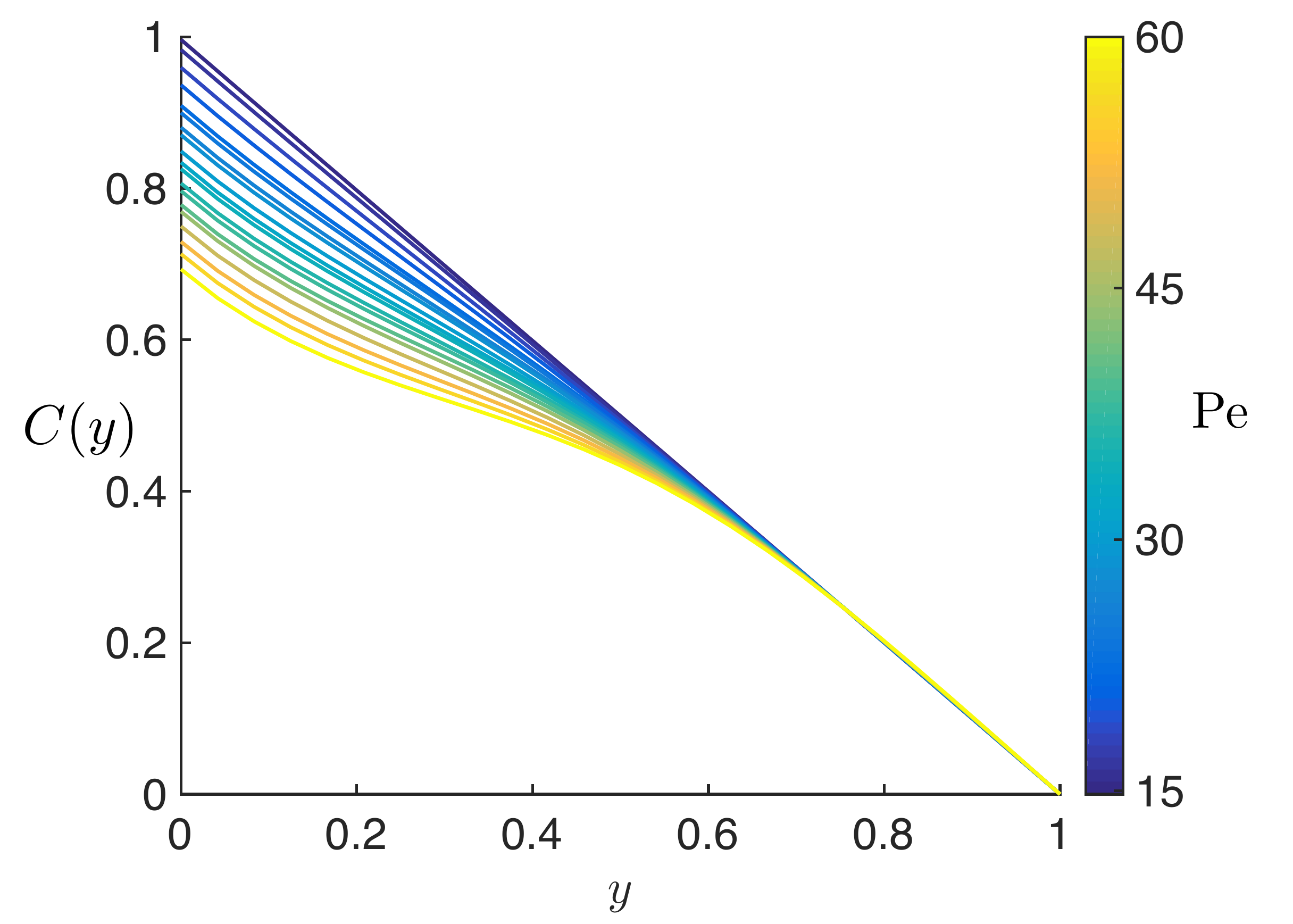}}
\end{tabular}
\caption{(a): Evolution     with the P\'eclet number, $\Pe$, of the mean concentrationwithin the channel $C_m$ (solid blue) and flow magnitude (as measured along the active wall by $\Lambda$ in Eq.~\eqref{eq:lambda}, dotted red) in a periodic uniform phoretic channel with $AM=1$ and $L=10\lambda_0\approx 24.4$ ($\lambda_\textrm{0}$ is the wavelength of the mode of optimal growth rate at the threshold $\Pe=\Pe_0$). For each value of $\Pe$,  20 different simulations are performed with  random initial  perturbations of the   steady state and the resulting standard deviation of both quantities is shown in shade. 
(b): Evolution with $\Pe$ of the $x$-averaged concentration  within the channel, $C(y)$,  for the same configuration.}\label{fig:meanC_Pe}
\end{center}
\end{figure}

\begin{figure*}[t]
\begin{center}
\includegraphics[width=.85\textwidth]{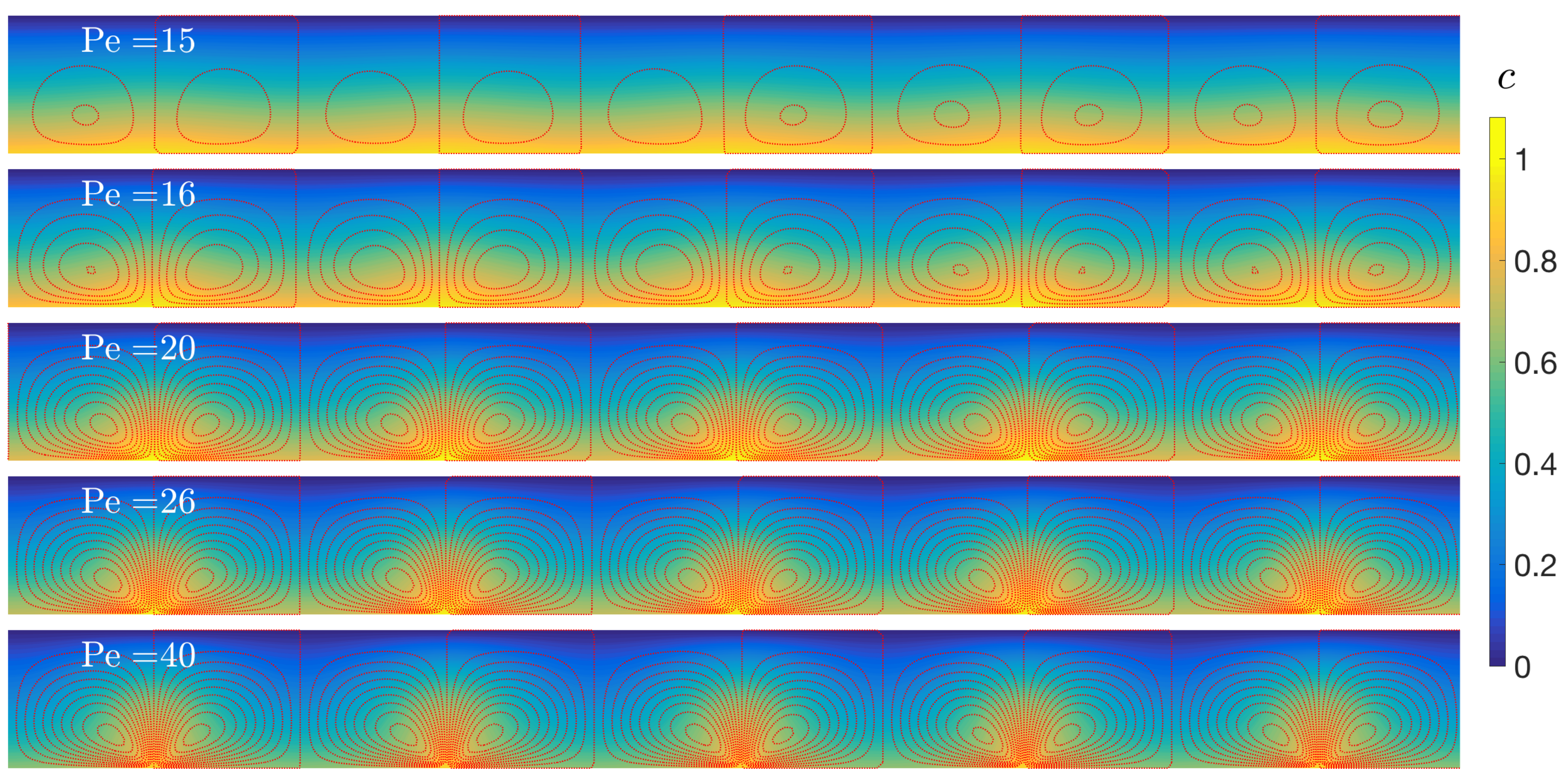}
\caption{Concentration (color) and streamlines (dotted lines) in a periodic uniform phoretic channel with $AM=1$, $L=10$ and increasing P\'eclet number, $\Pe$, for the final steady state solution associated with $5$ regular cells. The same scales are used for all panels for both concentration and streamfunction levels.}\label{fig:steady_state_Pe}
\end{center}
\end{figure*}

The dependence of the flow magnitude, $\Lambda$,  with the P\'eclet number, $\Pe$,   is plotted in  Figure~\ref{fig:bifurcation} for each steady state branch, while its average over several independent unsteady simulations is also shown on Figure~\ref{fig:meanC_Pe}(a). Increasing the value of  $\Pe$ beyond the instability threshold  results in an intensification, and saturation, of the periodic flow cells in response to the convective accumulation of solute concentration at definite regions along the active boundary (Figure~\ref{fig:steady_state_Pe}). As $\Pe$ is increased (in particular for $\Pe\geq 25$), a strong variability of the steady-state value of $\Lambda$ is observed between  simulations initiated from different random perturbations.

In turn, this enhanced phoretic flow for larger values of $\Pe$ profoundly modifies the solute distribution both along and across the channel. Indeed, as a result of the phoretic flow generated at the bottom active boundary, an upward (resp.~downward) convective transport of solute-rich (resp.~solute-depleted) fluid is observed in the regions of maximum (resp.~minimum) solute concentration at the boundary, resulting in a vertical convective transport of solute from the active to the passive boundary. The total chemical flux across the channel is imposed here by the fixed-flux chemical boundary condition in Eq.~\eqref{eq:bc2} along the active boundary. At steady state, the total solute flux across any horizontal surface is therefore independent of $y$ and it includes both a diffusive and convective contribution. The latter vanishes at the top and bottom boundaries (where $v=0$) but can be significant in the bulk of the channel. For large convective transport (large value of $\Pe$ and strong $\Lambda$), a reduction of the diffusive flux is therefore expected, which results in an overall reduction of the contrast between the solute concentration at the top (passive) and bottom (active) boundaries. The reference concentration of the upper wall is imposed here, and a net increase in convective transport  results therefore in a net reduction of the mean solute content within the channel, as quantified by the second measurements presented on Figure~\ref{fig:meanC_Pe}(a), namely the average $C_m$ of $c(x,y)$ over the entire channel. 

Further, the relative magnitude of diffusive and convective transport (and its impact on the concentration distribution) is not uniform across the channel due to the concentration of the flow cells driven by the active bottom boundary in the bottom half of the channel width. Specifically, the phoretic flow and resulting convective transport is greater in the bottom region (but away from the active wall where $v=0$), resulting in enhanced convective transport and reduced diffusive flux for $0.25\lesssim y\lesssim 0.5$. As a result, for increasing $\Pe$, the vertical gradient of horizontally-averaged concentration $C(y)$ is reduced in that region (Figure~\ref{fig:meanC_Pe}b). In contrast, within the top half of the channel and in the immediate vicinity of the bottom wall, the convective vertical flux of solute is almost negligible ($v\simeq 0$) and the vertical gradient of solute remains essentially unchanged.

It should be noted that the concentration distribution and flow field are specific to each steady state solution. With the present system exhibiting multistability, the final steady state observed numerically depends on the initial and periodic boundary condition imposed. The results presented in Figure~\ref{fig:meanC_Pe} are thus obtained for a set of 20 different random perturbations of the   base state, and their statistical mean and standard deviation are shown for each value of $\Pe$ considered. Nevertheless, we observe that the global quantities $\Lambda$ and $C_m$ depend only weakly on the precise configuration of the final steady state (i.e.~number of rotating cells), and the same holds for the vertical variations of $C(y)$, which confirms the generality of the results presented on Figure~\ref{fig:meanC_Pe}. 

\section{Conclusions}\label{sec:conclusions}

In this work, we demonstrated the emergence of spontaneous convective flows within a straight phoretic channel whose active walls drive a fluid flow in response to self-induced gradients of a chemical species. This mechanism,  which is similar to  a B\'enard-Marangoni instability, identifies therefore a  new route to the emergence of phoretic flows within a chemically-active micro-channel, in addition to asymmetric designs in chemical activity~\citep{michelin2019b} or wall geometry~\citep{michelin2015b}, similarly to the dual problem of   emergence of self-propulsion for phoretic colloids. 

A major difference between phoretic particles and channels  need to be emphasized. While chemical and phoretic asymmetry are observed to generate  net fluid transport (i.e.~self-propulsion of the colloid or net pumping above the active wall),   the phoretic instability which enables isotropic phoretic particles or active droplets to swim  cannot drive a net pumping flow through the straight channel. In essence, driving a net flow through the channel or self-propulsion of a colloid  requires a left-right symmetry breaking of the concentration distribution along the active wall which should be maintained by the resulting phoretic flow in order to obtain a self-sustained regime. This was possible around a colloidal particle due to  its surface curvature, which enables the formation of a solute-rich wake under the influence of advection. Such a mechanism  is  however not possible along an infinite flat wall.

Although no net flow is driven through the channel, the activity of the wall and the resulting phoretic instability allow for the development of steady and coherent convective cells that  profoundly modify the transport and distribution of solute within the micro-channel. In the present formulation, the total flux of chemical through the channel width is fixed by the wall catalytic activity, hence convective transport reduces the magnitude of the diffusive one and thus the concentration contrast of the channel with the constant level imposed by the passive wall. This effect   has the same origin as the enhancement of the thermal flux across a fluid gap undergoing a B\'enard-Marangoni instability when the temperature  levels of the lower and upper walls are prescribed.

Furthermore, the effect of the nonlinear advective coupling between chemical transport and phoretic flows could potentially drastically modify the saturated dynamics of an active channel with a (weak) design asymmetry (e.g. either non-symmetric geometry~\cite{michelin2015b} or chemical patterning of the wall~\cite{michelin2019b}), with a potential opportunity to significantly enhance the net pumping of such asymmetric designs. A parallel could indeed be drawn here with the self-propulsion of an almost isotropic spherical particle (e.g. an active spherical particle with a small inert patch). The swimming velocity of such particles is small in the diffusive limit ($\Pe=0$), typically scaling with the degree of asymmetry of the system; yet, for finite $\Pe$, a finite $O(1)$ swimming velocity is achieved due to the nonlinear coupling of solute transport and phoretic flows~\cite{michelin2014}.

The  results in our paper were obtained within the simplified chemical framework of a fixed rate of release/consumption of solute at the active boundary and a steady and uniform concentration at the passive wall. Our analysis could be  generalized to more complex chemical kinetics, leading to additional dimensionless characteristics of the problem, such as a reaction-to-diffusion ratio. One such example is a first-order reaction at the active wall where the rate of consumption of solute is proportional to its local concentration, thereby allowing for an additional self-saturation of the chemical reaction when diffusion is not sufficiently fast to replenish the solute content near the active wall. This particular case has been considered in the case of self-propulsion of active colloids~\cite{cordova2008,michelin2014}. \change{Similarly, the approach in our paper could be generalized also to analyze a different combination of activity and mobility on both of the channel walls (e.g., one wall with activity but no mobility and the other with mobility only, or both walls with the two properties). }Although quantitative results would then depend on the exact physico-chemical model for the surface chemistry \change{and the relative activity and mobility of the two walls}, the  emergence of spontaneous convective phoretic flows and resulting modification of the convective transport reported here are expected to hold generically. \change{Finally, we note that the present analysis is based on short-ranged interactions of the solute molecules with the channel walls and, as a result, the flow forcing is expressed as a slip velocity of the concentration gradient at the wall. If the interaction thickness is no longer negligible in comparison with the  channel width, the present approach could  be generalized by directly including  the solute-wall interaction forces in the momentum balance~\cite{anderson1989,michelin2014}.}

\section*{Conflicts of interest}
There are no conflicts to declare.

\section*{Acknowledgements}
This project has received funding from the European Research Council (ERC) under the European Union's Horizon 2020 research and innovation programme  (Grant agreements  714027 to SM and 682754 to EL) as well as the Engineering and Physical Sciences Research Council (EPSRC, Grant EP/L020564/1 to DTP).

\end{document}